\documentclass[12pt]{article}

\usepackage{amsmath,amssymb}
\usepackage{cite}

\newcommand{\dd}{\partial}
\newcommand{\m}{\mu}
\newcommand{\ls}{\left(}
\newcommand{\rs}{\right)}
\newcommand{\al}{\alpha}
\newcommand{\be}{\beta}
\newcommand{\ff}{\varphi}
\newcommand{\te}{\theta}
\newcommand{\ga}{\gamma}
\newcommand{\sh}{\sinh}
\newcommand{\ch}{\cosh}

\newcommand{\disn}[2]{$$\displaylines{\refstepcounter{equation}%
            \label{#1}\hskip 1em minus 1em #2\hfilneg}$$}
\newcommand{\nom}{\hfil\hskip 1em minus 1em (\theequation)}
\newcommand{\no}{\hfil \hskip 1em minus 1em\phantom{(\theequation)}%
            \hfilneg\cr\hfilneg\hskip 1em minus 1em\hfil}

\begin{document}

\title{When does the Hawking into Unruh mapping\\ for global embeddings work?}
\author{S.A.~Paston\thanks{E-mail: paston@pobox.spbu.ru}\\
{\it Saint Petersburg State University, St.-Petersburg, Russia}
}
\date{\vskip 15mm}
\maketitle

\begin{abstract}
We discuss for which smooth global embeddings of a metric into a Minkowskian spacetime the Hawking into Unruh mapping takes place.
There is a series of examples of global embeddings into the Minkowskian spacetime (GEMS) with such mapping for physically interesting metrics.
These examples use Fronsdal-type embeddings for which timelines are hyperbolas.
In the present work we show that for some new embeddings (non Fronsdal-type) of the Schwarzschild
and Reissner-Nordstr\"om metrics there is no mapping.
We give also the examples of hyperbolic and non hyperbolic type embeddings for the de Sitter metric
for which there is no mapping.
For the Minkowski metric where there is no Hawking radiation we consider a non trivial embedding with hyperbolic timelines, hence in the ambient space the Unruh effect takes place, and it follows that there is no mapping too.
The considered examples show that the meaning of the Hawking into Unruh mapping for global embeddings remains still
insufficiently clear and requires further investigations.
\end{abstract}

\newpage

\section{Introduction}
The Hawking radiation \cite{hawking75} should be observed as a quantum effect of the emission of particles by a black hole. It is caused by a non trivial metric of a Riemannian space which corresponds to the black hole and is considered as a classical one.
This radiation can be described if we consider the vector of state, which is vacuum from the point of view of field operators at the initial surface, and we decompose it over the basis of the states corresponding to field operators at the future finite surface which includes the infinitely distant surface and the black hole horizon.
The radiation spectrum is thermal with the temperature $T_H=1/(8\pi m)$ for the Schwarzschild black hole with the mass $m$.

The Unruh effect \cite{unruh} predicts that while an inertial observer in a Minkowski space sees a vacuum state, an observer moving in the same state with a constant acceleration $w$ will detect a thermal radiation with the temperature $T_U=w/(2\pi)$.
This effect is usually described using an Unruh-DeWitt detector moving over some trajectory and interacting with a quantum field existing in the spacetime. A similar effect is also found for other accelerated motions, but the radiation spectrum is no more thermal \cite{letaw81}.

For the Schwarzschild metric with the interval
\disn{0.1}{
ds^2=\ls 1-\frac{2m}{r}\rs dt^2-\frac{dr^2}{1-\frac{2m}{r}}-r^2\ls d\te^2+\sin^2\te\, d\ff^2\rs,
\nom}
it has been  found out in \cite{deserlev99} that for a Fronsdal's isometric embedding \cite{frons} in a flat 6-dimensional space with the $(+-----)$ signature
\disn{1}{
r>2m: \qquad
y^0=\tau\sqrt{1-\frac{2m}{r}}\,\sh\ls\frac{t}{\tau}\rs,\qquad
y^1=\pm\tau\sqrt{1-\frac{2m}{r}}\,\ch\ls\frac{t}{\tau}\rs,\no
r<2m: \qquad
y^0=\pm\tau\sqrt{\frac{2m}{r}-1}\,\ch\ls\frac{t}{\tau}\rs,\qquad
y^1=\tau\sqrt{\frac{2m}{r}-1}\,\sh\ls\frac{t}{\tau}\rs,\no
y^2=\int dr\sqrt{\frac{2m}{r-2m}\ls 1-\frac{m\tau^2}{2r^3}\rs},\no
y^3=r\,\cos\te,\quad
y^4=r\,\sin\te\,\cos\ff,\quad
y^5=r\,\sin\te\,\sin\ff
\nom}
the Hawking radiation can be interpreted as the Unruh effect (the Hawking into Unruh mapping, also named GEMS (global embedding Minkowskian spacetime) approach). The reason of such interpretation is the fact that an observer being stationary at fixed $r$ in the Rimannian space moves with a constant acceleration in the ambient space, and a thermal radiation occurs from both points of view and has  the same temperature.
It is very interesting that the temperatures are equal for the unique value of the embedding parameter $\tau=4m$ at which the component $y^2$ in (\ref{1}) does not contain singularities and the embedding smoothly covers the event horizon.

The obtained result caused a number of works where the Hawking into Unruh mapping was tested in many other cases.
The corresponding results were obtained for other types of black holes (including the black holes with an electric charge, with a cosmological constant and with the dimension other than 4), black strings, wormholes, as well as de Sitter and anti de Sitter spaces,
see \cite{deserlev99,kim00,lemos,arXiv:1012.5709,hep-th/0103036,gr-qc/0303059,arXiv:1311.0592} and cited herein.
A detector moving over a circumference \cite{gr-qc/0409107} or in a free fall \cite{arXiv:0805.1876} near the black hole has also been discussed. In these cases the radiation spectrum is no more exactly thermal.
It has been shown in \cite{arXiv:0901.0466} that the coincidence is still valid for the corrections to the Hawking and Unruh temperatures.
It has been noted in \cite{Banerjee} that the Hawking into Unruh mapping for black holes takes place also if we embed the $(t-r)$ sector of the Riemannian space only.
Note that in all cases the Fronsdal-type embeddings were used in which the timelines are hyperbolas in the ambient space.

The Hawking effect results from the interaction of the detector with a quantum field defined in the Riemannian space, for example with a scalar massless field satisfying the invariant equation \cite{hawking75}
\disn{2}{
\nabla_\m \nabla^\m \ff(x)=0.
\nom}
The corresponding Unruh effect results from the interaction of the detector with a quantum field defined in the flat ambient space and satisfying the equation
\disn{3}{
\dd_a \dd^a \ff(y)=0
\nom}
(here $\m=0,1,2,3$, and $a$ ranges over the dimension of the ambient space).
The solutions of the equations (\ref{2}) can be extended up to the solutions of the equations (\ref{3}) which satisfy an additional condition that the two first derivatives in the "transverse" directions are equal to zero
(the link between the equations (\ref{2}),(\ref{3}) is used, for example, in the study of the relation between thermodynamics and horizons in \cite{padmanabhan02}).
This is why the existence of the Hawking into Unruh mapping can mean a close relationship between quantum theories in a Riemannian space and in its flat ambient space.
Such relationship supports the suggestion that the gravity, usually described in the framework of General Relativities as a result of the Riemannian space curvature, can be considered as an embedding theory where the 4-dimensional spacetime is a surface in a flat space with higher dimension. The quantum theory can be formulated just in this flat space, and such approach can be useful for the construction of a correct quantum gravitation theory.
Various modifications of the embedding theory were developed in \cite{regge,deser,pavsic85let,tapia,davkar,statja18,statja24} and other works.
In the work \cite{statja25} the gravity is formulated as a field theory in a flat ambient space, where the matter fields are also described by some functions in this ambient space.

The aim of the present work is to discuss the following problem: is the Hawking into Unruh mapping valid for any smooth (smoothly covering the horizon as well) embedding of some metric? If the existence of this mapping is not a simple accident, but it takes place for some special embeddings only, then we have to understand why the mapping works just only for them. However, we still cannot do that completely.

In Sections 2 and 3 we discuss the recently found embeddings for the Schwarzschild metric and the Reissner-Nordstr\"om metric respectively.
For these embeddings the timelines appear to be not hyperbolas,
i.e. they are not Fronsdal-type embeddings,
and we demonstrate the absence of the Hawking into Unruh mapping for them.
In Section 4 we suggest a non Fronsdal-type embedding, as well as a Fronsdal-type embedding for the de Sitter metric,
with no mapping for them both.
Finally, in Section 5 we consider simple Fronsdal-type embeddings for the Minkowski metric, for which there is no mapping as well.
Hence we see the examples of the absence of the Hawking into Unruh mapping for both non Fronsdal-type and Fronsdal-type embeddings.

\section{Schwarzschild metric embeddings without mapping}
At the time when the main article \cite{deserlev99} on the subject was written, only three embeddings of the Schwarzschild metric in a flat 6-dimensional space were known:
those of Kasner \cite{kasner3}, Fronsdal \cite{frons} and Fujitani-Ikeda-Matsumoto \cite{fudjitani}.
The embeddings \cite{kasner3} and \cite{fudjitani} cannot be smoothly extended under the horizon,
so only the Fronsdal's embedding (\ref{1}) was known as covering smoothly the horizon (at $\tau=4m$). This embedding has hyperbolic timelines.
Three other 6-dimensional Schwarzschild metric embeddings which smoothly cover the horizon were found later in \cite{davidson,statja27}.
We can demonstrate the absence of the Hawking into Unruh mapping for these isometric embeddings.

Three new 6-dimensional isometric embeddings of the Schwarzschild metric are (see the details in \cite{statja27}):
the cubic embedding
 \disn{4}{
\begin{array}{l}
\displaystyle y^0=\frac{\xi^2}{6}t'^3+\ls 1-\frac{m}{r}\rs t'+u(r),\\[1em]
\displaystyle y^1=\frac{\xi^2}{6}t'^3-\frac{m}{r}t'+u(r),\\[1em]
\displaystyle y^2=\frac{\xi}{2}t'^2+\frac{1}{2\xi}\ls 1-\frac{2m}{r}\rs,
\end{array}
\nom}
the Davidson-Paz embedding
\disn{5}{
\begin{array}{l}
\displaystyle y^0=\frac{m}{\be\sqrt{r_c r}}\ls e^{\be t'+u(r)}-\frac{r-r_c}{2m}\,e^{-\be t'-u(r)}\rs,\\[1em]
\displaystyle y^1=\frac{m}{\be\sqrt{r_c r}}\ls e^{\be t'+u(r)}+\frac{r-r_c}{2m}\,e^{-\be t'-u(r)}\rs,\\[1em]
\displaystyle y^2=\hat\ga t',
\end{array}
\nom}
and the asymptotically flat embedding
 \disn{6}{
\begin{array}{l}
\displaystyle y^0=t',\\[0em]
\displaystyle y^1=\frac{(6m)^{3/2}}{\sqrt{r}\quad}\,\sin\ls\frac{t'}{3^{3/2}2m}-\sqrt{\frac{2m}{r}}\,\bigg( 1+\frac{r}{6m}\bigg)^{3/2}\,\rs,\\[1.2em]
\displaystyle y^2=\frac{(6m)^{3/2}}{\sqrt{r}\quad}\,\cos\ls\frac{t'}{3^{3/2}2m}-\sqrt{\frac{2m}{r}}\,\bigg( 1+\frac{r}{6m}\bigg)^{3/2}\,\rs.
\end{array}
\nom}
In these formulas $t'$ is the time for some falling coordinates; $\xi,\be,\hat\ga,r_c$ are the parameters of the embeddings, and the components $y^3,y^4,y^5$ together with the signature of the ambient space coincide with those in the embedding (\ref{1}).
The timelines for all these embeddings are not hyperbolic. They represent the stationary motions, and the effect analogous to the Unruh effect for them was studied in \cite{letaw81}.

The timelines of the embedding (\ref{4}) correspond to the cusped motion, and for this motion the spectrum is exactly calculated \cite{letaw81}
 \disn{7}{
S(E)=\frac{E^2}{8\pi^2\sqrt{3}\,\xi^2}\ls 1-\frac{2m}{r}\rs^2
\exp\ls -\frac{\sqrt{12}}{\xi^2}\ls 1-\frac{2m}{r}\rs E\rs,
\nom}
and is not thermal.

For the motion along the timelines of the embedding (\ref{5}) the spectrum was studied numerically in \cite{letaw81}, it was also analyzed in the recent work \cite{abdolrahimi}. This spectrum is not thermal as well.

The timelines of the embedding (\ref{6}) correspond to the circular motion of the detector. The effect analogous to the Unruh effect for this case was extensively studied in a number of works, see the references in \cite{abdolrahimi}. There is an opinion that it can be
 experimentally observed \cite{Akhmedov}.
The spectrum in this case is not thermal as well. Moreover, it is easy to see from (\ref{6}) that at $r\to\infty$ the radius of the circumference tends to zero (at a fixed angular speed),
and the detector rests in a selected coordinate system of the ambient space, so there must be no Unruh effect in this limit.
This situation is of course due to the fact that the embedding (\ref{6}) is asymptotically flat, i.e. the corresponding surface at $r\to\infty$ tends to a plane (unlike the embeddings (\ref{1}),(\ref{4}),(\ref{5})). This is the embedding which seems to be the more natural from the point of view of the embedding theory, see the discussion in \cite{statja27}.

The accelerations for the detector moving over the timelines in the case of the embeddings (\ref{4})-(\ref{6}) were calculated in \cite{statja32}.
Hence we conclude that for all the three new 6-dimensional embeddings of the Schwarzschild metric there is no Hawking into Unruh mapping.

\section{Reissner-Nordstr\"om metric embeddings\\ without mapping}
For the Reissner-Nordstr\"om metric
 \disn{7.1}{
ds^2=\left(1-\frac{2m}{r}+\frac{q^2}{r^2}\right)dt^2-\left(1-\frac{2m}{r}+\frac{q^2}{r^2}\right)^{-1}dr^2
-r^2 \ls d\te^2+\sin^2\te\, d\ff^2\rs,
\nom}
which describes a black hole with the charge $q$,
an embedding
in a flat 7-dimensional space with the $(++-----)$ signature was proposed in the work \cite{deserlev99}:
 \disn{7.2}{
r>r_+:\qquad
y^0=\be^{-1}\sqrt{1-\frac{2m}{r}+\frac{q^2}{r^2}}\;\sh(\be t),\quad
y^1=\pm \be^{-1}\sqrt{1-\frac{2m}{r}+\frac{q^2}{r^2}}\;\ch(\be t),\no
r_-<r<r_+:\;\;
y^0=\pm \be^{-1}\sqrt{\frac{2m}{r}-1-\frac{q^2}{r^2}}\;\ch(\be t),\;\;
y^1=\be^{-1}\sqrt{\frac{2m}{r}-1-\frac{q^2}{r^2}}\;\sh(\be t),\no
y^2=r\cos\te,\qquad y^3=r\sin\te\,\cos\ff,\qquad y^4=r\sin\te\,\sin\ff,\no
y^5=\int\! dr\, \sqrt{\frac{r^2 (r_++r_-)+r^2_+(r+r_+)}{r^2 (r-r_-)}},\qquad
y^6=\frac{2\sqrt{r_+^5 r_-}}{(r_+-r_-)r},
\nom}
where $\be=(r_+-r_-)/(2r_+^2)$ and $r_\pm$ are the roots of the equation $g_{00}(r)=0$.
This embedding smoothly covers the external horizon $r=r_+$ and
the Hawking into Unruh mapping was found for it in the work \cite{deserlev99}.
Note that this is  a hyperbolic type embedding and
that it cannot be smoothly continued tothe  $r<r_-$ region,
hence this embedding is not global in the full sense of the word.

Recently three new embeddings of the Reissner-Nordstr\"om metric were found in the work \cite{statja30}.
These embeddings are smooth at all $r>0$, i.e.  they smoothly cover  both horizons and they are really global.
These three embeddings in a flat 6-dimensional space with the $(++----)$ signature
are (see the details in \cite{statja30}):
the cubic embedding
\begin{eqnarray}\label{7.3}
&& y^0 = \frac{2m^3}{q^4}t'^2-\frac{q^4}{8m^3}\ls 1-\frac{2m}{r}+\frac{q^2}{r^2}\rs,
\nonumber\\
&& y^1 = \frac{4m^6}{3q^8}t'^3-\frac{1}{4}\ls 1-\frac{2m}{r}+\frac{q^2}{r^2}\rs t'-t'+u\!\ls\frac{2mr}{q^2}\rs,
\nonumber\\
&& y^2 = \frac{4m^6}{3q^8}t'^3-\frac{1}{4}\ls 1-\frac{2m}{r}+\frac{q^2}{r^2}\rs t'+t'+u\!\ls\frac{2mr}{q^2}\rs,
\end{eqnarray}
the exponential embedding
\begin{eqnarray}\label{7.4}
&& y^0 = \ga t',\nonumber\\
&& y^1 = \frac{1}{2\beta} \left(e^{-\be t' - u\ls 2mr/q^2\rs} -
 \ls 1-\frac{2m}{r}+\frac{q^2}{r^2}-\ga^2 \rs  e^{\beta t' + u\ls 2mr/q^2\rs}
 \right),\nonumber\\
&& y^2 = \frac{1}{2\beta} \left(e^{-\be t' - u\ls 2mr/q^2\rs} +
 \ls 1-\frac{2m}{r}+\frac{q^2}{r^2}-\ga^2 \rs  e^{\beta t' + u\ls 2mr/q^2\rs}
 \right),
\end{eqnarray}
and the spiral embedding
\begin{eqnarray}\label{7.5}
&& y^0 = \frac{\sqrt{(mr-q^2)^2+b^2 r^2}}{\al q r}\sin\ls\alpha t' + u\!\ls\frac{2mr}{q^2}\rs\rs,\nonumber\\
&& y^1 = \frac{\sqrt{(mr-q^2)^2+b^2 r^2}}{\al q r}\cos\ls\alpha t' + u\!\ls\frac{2mr}{q^2}\rs\rs,\nonumber\\
&& y^2 = \frac{\sqrt{b^2+m^2-q^2}}{q}\; t'.
\end{eqnarray}
In these formulas $t'$ is the time for some falling coordinates; $\ga,\be,\al,b$
are the parameters of the embeddings, and the components $y^3,y^4,y^5$ coincide with those in the embedding (\ref{1}).

It is easy to see that the structure of the embeddings (\ref{7.3})-(\ref{7.5})
is very similar to the embeddings (\ref{4})-(\ref{6})  discussed in the Section 2.
The timelines have the analogical form and are not hyperbolic.
Hence for the three new globally smooth embeddings of the Reissner-Nordstr\"om metric found in \cite{statja30}
we can also conclude that there is no Hawking into Unruh mapping.

\section{De Sitter metric embeddings without mapping}
The mapping for the de Sitter space was also given in \cite{deserlev99}.
It is however not complicated to construct an embedding for the de Sitter metric which is not a standard hyperboloid.
For example, it is always possible to bend isometrically the flat space containing this hyperboloid.
The result is an embedding in the flat space with the $(+-----)$ signature, under the form
 \disn{8}{
\begin{array}{l}
y^0=R\, \sinh(\sinh t),\\
y^1=R\, \cosh(\sinh t),\\
y^2=R\, \cosh t\,\cos\chi,\\
y^3=R\, \cosh t\,\sin\chi\,\cos\te,\\
y^4=R\, \cosh t\,\sin\chi\,\sin\te\,\cos\ff,\\
y^5=R\, \cosh t\,\sin\chi\,\sin\te\,\sin\ff,
\end{array}
\nom}
for which the timelines are no more hyperbolas (and even not the stationary trajectories),
hence the spectrum corresponding to the motion along it will not be thermal, and there will be no mapping.

It might seem that the hyperbolicity of an embedding is a mapping criterion, but this is not so. If we use the method suggested in \cite{willison1302} for the construction of a nonstandard embedding for the anti de Sitter space, we can construct one more non trivial embedding for the de Sitter space:
 \disn{8.1}{
\begin{array}{l}
y^0=R\al^{-1}\cos\chi\,\sinh\al t,\\
y^1=R\al^{-1}\cos\chi\,\cosh\al t,\\
y^2=R\, \sin\chi\,\cos\te,\\
y^3=R\, \sin\chi\,\sin\te\,\cos\ff,\\
y^4=R\, \sin\chi\,\sin\te\,\sin\ff,\\
y^5=R\sqrt{1-\al^{-2}}\cos\chi,
\end{array}
\nom}
which transforms at $\al=1$ into a standard one (the signature is again $(+-----)$).
Here the timelines are hyperbolas, however the corresponding acceleration and consequently the Unruh temperature depend on the arbitrary parameter $\al$, while the metric does not depend on it. Hence, although both spectra (of both radiations, observed by the detector either in the internal geometry or from the point of view of the motion in the ambient space) are thermal,
there will be no mapping for the given embedding if $|\al|<1$. The same result can be formulated \cite{willison_pv} for the nonstandard embedding of the anti de Sitter space suggested in \cite{willison1302} (a mapping for the standard embedding was also studied in \cite{deserlev99}).

\section{Minkowski metric embedding without mapping\\ and some discussion}
There exists a more simple example of the absence of mapping for an embedding with hyperbolic timelines.
Consider a surface defined by the embedding function
 \disn{9}{
\begin{array}{l}
y^0=\al^{-1} \sinh \al t,\\
y^1=\al^{-1} \cosh \al t,\\
y^2=r\cos\te,\\
y^3=r\sin\te\cos\ff,\\
y^4=r\sin\te\sin\ff
\end{array}
\nom}
with the ambient space signature $(+----)$.
One can easily see that it corresponds to a metric matching the metric of the Minkowski space, i.e.~(\ref{9}) defines a non trivial isometric embedding of the Minkowski space in a flat 5-dimensional ambient space.
If an Unruh-DeWitt detector moves over the timeline, then from the point of view of the internal geometry it will not observe any radiation, because it moves along an inertial trajectory in a 4-dimensional Minkowski space. However in the ambient space the timeline is a hyperbola, the detector is uniformly accelerated, and a standard Unruh effect should be observed.
Hence in the given example there is no mapping as well.

As we mentioned above the Hawking into Unruh mapping takes place usually
for the embeddings with hyperbolic timelines.
We can partly understand why the mapping can take place just  for the embeddings of such type.
In this case the embedding function depends  on time as
 \disn{10}{
\begin{array}{l}
y^0=f(r)\sinh \al t,\\
y^1=f(r)\cosh \al t.\\
\end{array}
\nom}
Therefore, on the one hand, we obtain $g_{00}(r)=\al^2f(r)^2$, and on the other hand
the acceleration of the observer which moves  in the ambient space along the timelines is $w(r)=1/f(r)$.
Hence for the Unruh temperature $T_U$ the condition
 \disn{11}{
T_U(r)\sqrt{g_{00}(r)}=const
\nom}
will be satisfied  automatically. This condition  coincide with the known Tolman law for a locally measured temperature of the Hawking radiation, therefore the dependences of Hawking and Unruh temperatures on $r$ will be identical.
However it is still unclear what determines the fact that the Hawking and Unruh temperatures coincide or not at the infinity.

The examples considered in this work show that
the Hawking into Unruh mapping should be used with great care
for obtaining information about thermodynamic properties of spaces with horizons,
and the question where the Hawking into Unruh mapping does work or not and what is its nature needs further investigation.

\vskip 0.5em
{\bf Acknowledgments}.
The author is grateful to A.A.~Sheykin for useful discussions and to S.~Willison
for a valuable remark and a relevant reference.
The author acknowledge Saint-Petersburg State University for a research grant 11.38.660.2013.

\providecommand{\href}[2]{#2}\begingroup\raggedright
\endgroup


\end{document}